\documentclass[prl,superscriptaddress,floatfix,citeautoscript,longbibliography,twocolumn]{revtex4-2}
\usepackage{amsmath}
\usepackage{xcolor}
\usepackage{braket}
\usepackage{hyperref}
\usepackage{float}
\usepackage{graphicx} 
\usepackage{hyperref}
\usepackage{float}
\usepackage{graphicx}
\usepackage{listingsutf8}
\usepackage{tcolorbox}
\usepackage{colortbl}
\usepackage{multirow}

\newcommand{\lcpq}{Laboratoire de Chimie et Physique Quantiques, CNRS, Universit\'e de Toulouse, 118 Route de Narbonne, F-31062 Toulouse, France}
\newcommand{\bologna}{Universit\`a di Bologna, 40126 Bologna, Italy}
\newcommand{\etsf}{European Theoretical Spectroscopy Facility (ETSF)}

\newcommand{\br}{\mathbf{r}}

\begin{document}

\title{Quantum chemistry for solids made simple on the Clifford torus}

\author{Amer Alrakik}
\affiliation{\lcpq}
\affiliation{\etsf}

\author{Gian Luigi Bendazzoli}
\affiliation{\bologna}

\author{Stefano Evangelisti}
\affiliation{\lcpq}

\author{J.~Arjan Berger}
\affiliation{\lcpq}
\affiliation{\etsf}
\email{arjan.berger@irsamc.ups-tlse.fr}

\begin{abstract}
We present a general theory to treat periodic solids with quantum-chemistry methods.
It relies on two main developments: 1) the modeling of a solid as a Clifford torus which is a torus that is both periodic and flat and 
2) the introduction of a periodic gaussian basis set that is compatible with the topology of the Clifford torus.
We illustrate our approach by calculating the ground-state energy of a periodic chain of hydrogen atoms within both Hartree-Fock and coupled cluster theory.
We demonstrate that our approach yields the correct ground-state energy in the thermodynamic limit by comparing it to the ground-state energy of a ring of hydrogen atoms in the same limit.
Since equivalent ring-like calculations for three-dimensional solids are impossible, our approach is an excellent alternative to perform quantum-chemistry calculations of solids.
Our Clifford formalism can be seamlessly combined with current implementations of quantum-chemistry methods designed for atoms and molecules to make them applicable to solids.
\end{abstract}

\maketitle
%
%
Quantum chemistry methods, such as configuration interaction and  coupled-cluster theory, are among the most accurate theories to describe the electron correlation in atoms and molecules.
Although several approaches exist to extend these methods to periodic solids~~\cite{Sun_1996,Ayala_2001,Pisani_2005,Hirata_2004,Katagiri_2005,Marsman_2009,Grueneis_2011,Pisani_2012,Booth_2013,Shephard_2012,McClain_2017,Wang_2020,Erba_2023}, quantum-chemistry calculations of solids are still far from routine.
Quantum-chemistry calculations of periodic systems have been hampered by two issues:
1) The lack of an efficient representation of a periodic system described by a Hamiltonian containing an explicit electron-electron interaction.
In particular, modelling a solid as an infinite system leads to divergencies in the theory~\cite{Sun_2017}.
2) The lack of a periodic basis that can accurately describe the sharp oscillating features of the wave function close to the nuclei.
Let us elaborate on these points in the following.

Due to the explicit treatment of the electron-electron interaction in quantum-chemistry methods,
the corresponding Hamiltonians do not obey translational invariance with respect to the translation of a single electron by a lattice vector.
As a consequence the translational symmetry of the crystal cannot be exploited to limit calculations to the unit cell of the crystal.
This is in contrast, for example, to Kohn-Sham density-functional theory, in which there is no explicit electron-electron interaction in the Hamiltonian.
Periodic systems described by Hamiltonians with an explicit electron-electron interaction could be represented by a periodic supercell containing several unit cells.
In one dimension (1D) such a supercell would correspond to a ring, or 1-torus, in two dimensions (2D) to a 2-torus, and in three dimensions (3D) to a 3-torus~\footnote{Real systems are never strictly one- or two-dimensional since the electronic density always extends over three dimensions.
However, they can be periodic in one, two or three dimensions and thus be described by a quasi-one-dimensional torus, a quasi-two-dimensional torus and a quasi-three-dimensional torus, respectively. The latter cannot be represented in a three-dimensional Euclidean space.
}.
The infinite periodic solid is then recovered by extrapolating to the thermodynamic limit.
A 1-torus and, although in a cumbersome way, even a 2-torus could, in principle, be represented within a three-dimensional Euclidean space.
However, this is strictly impossible for a 3-torus.

Instead, here we propose to represent the periodic supercell by a Clifford torus.
Just like an ordinary torus, a Clifford torus has no boundaries but, unlike an ordinary torus, it is flat, i.e., it has zero Gaussian curvature everywhere.
It can be obtained by joining the opposite edges of a line (1D), a rectangle (2D) or a rectangular parallelepiped (3D) \textit{without deformation}, i.e., without bending, stretching or twisting.
In order for a Clifford torus to be both flat and without boundaries it is embedded in a higher-dimensional Euclidean space. 
For example, a 1D Clifford torus is embedded in a 2D Euclidean space since it is topologically equivalent to a ring.
For 2D and 3D Clifford tori, the embedding spaces are in 4D and 6D, respectively.
The distance between two points on the Clifford torus is the length of the shortest path between the two points in the embedding space of the Clifford torus.
This naturally leads to a renormalization of the Coulomb potential which describes the interaction between two charged particles and inversely depends on the distance between the two particles~\cite{Aragao_2019,Tavernier_2020,Tavernier_2021,Escobar_2021,Alves_2021,Evangelisti_2022,Alrakik_2023,Escobar_2024}.
We note that our formalism is not related to Clifford algebra~\cite{Su_2024}

The most widely used basis functions in quantum-chemistry calculations are gaussian-type orbitals.
Gaussian-type orbitals, or gaussians, were introduced 75 years ago by Boys~\cite{Boys_1950} and revolutionized the electronic-structure calculations of atoms and molecules since the calculation of the corresponding one- and two-electron integrals could be performed in a simple and efficient manner~\cite{helgaker2013molecular}.
This is mainly due to the gaussian product rule according to which the product of two gaussian functions can be expressed as a single gaussian function.
Unfortunately, since gaussians are not periodic, they cannot be used straightforwardly in the calculation of periodic systems.
Instead, plane waves are periodic and could, in principle, be used as a basis in calculations of periodic systems.
Unlike gaussians, plane waves have difficulties to describe the sharp oscillating features of the wave function close to the nuclei
and typically require effective core potentials to keep calculations tractable.
However, this reduces the accuracy of calculations with a plane-wave basis.

Instead, here we propose a new type of periodic basis function, the Clifford gaussian, that has all the benefits of a regular gaussian while also being periodic.
In particular, we will demonstrate the existence a product rule for Clifford gaussian functions, i.e., the product of two Clifford gaussian functions can be expressed as a single Clifford gaussian function.
Moreover, we show that Clifford gaussians are naturally compatible with the topology of the Clifford torus.

In summary, with respect to quantum-chemistry calculation of molecules, our approach for solids only requires two modifications:
1) replace the distance in the Coulomb interaction by a renormalized distance.
2) replace regular non-periodic gaussians by periodic Clifford gaussians.
The properties of the solid are then recovered by extrapolating the results obtained for Clifford supercells of different sizes to the thermodynamic limit.
The crucial advantage of our approach is that these modifications only concern the calculation of the one- and two-electron integrals and the inter-nuclear repulsion but are completely \emph{independent} of the quantum-chemistry method used in the calculation.
Therefore, once these modifications have been applied, any quantum-chemistry method (Hartree-Fock, coupled cluster, etc,) can be used to perform calculations on periodic solids.
In essence, our approach reduces the calculation of a solid to several calculations of (big) molecules followed by an extrapolation to the thermodynamic limit.
The smoothness and efficiency of the extrapolation are guaranteed by the topology of the Clifford torus because it has no boundaries.
Moreover, by using the example of ground-state energy of a chain of hydrogen atoms, we will show that in the thermodynamic limit the results obtained within our approach are numerically equivalent to those obtained for the ring.
%
%

\begin{figure}[t]
    \centering
    \includegraphics[width=1.0\linewidth]{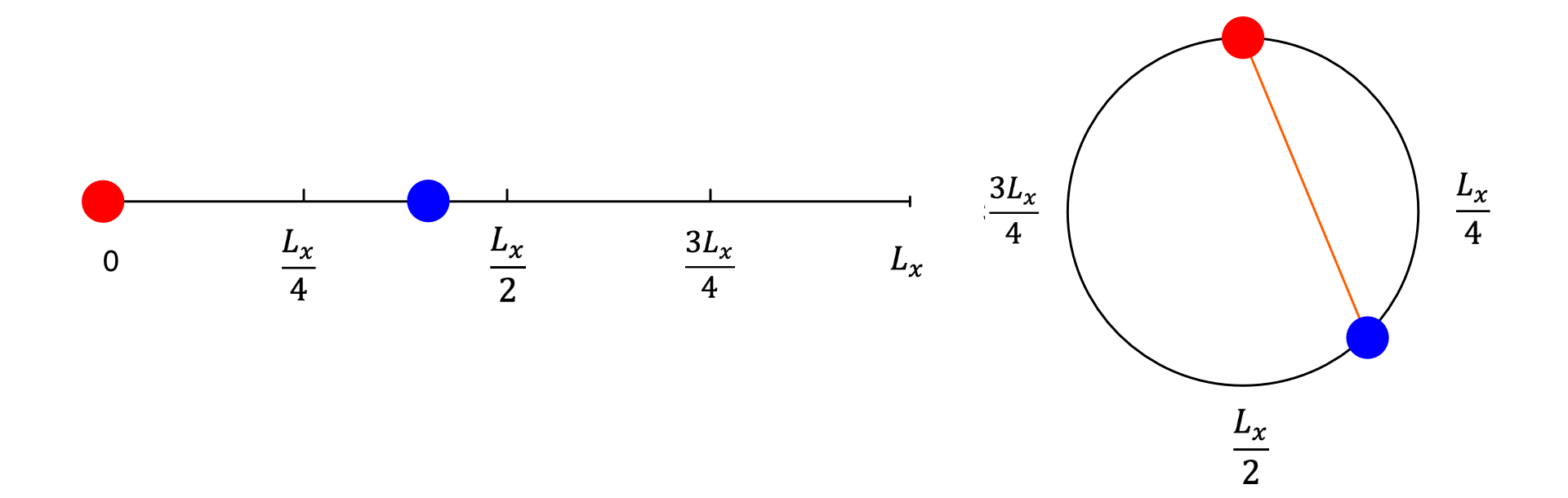}
    \caption{The Euclidean distance between two points in a one-dimensional Clifford supercell of length $L_x$. 
    Left panel: two points in a one-dimensional Clifford supercell. Right Panel : The Euclidean distance between these two points represented on a ring that is topologically equivalent to the Clifford torus.}
    \label{Distance}
\end{figure}

We represent a periodic solid by a supercell consisting of several unit cells.
We apply periodic boundary conditions by imposing the supercell to have a Clifford topology, i.e., opposite sides of the supercell are joined without deformation of the supercell.
The Hamiltonian corresponding to a supercell with the Clifford topology is given by (in Hartree atomic units)
\begin{align}
\label{Eq:Hamiltonian}
\hat{H} &= -\frac12\sum_{i=1}^{N_e} \nabla^2_{\br_i} - \sum_{A=1}^{N_n} \sum_{i=1}^{N_e} \frac{Z_A}{|\br_i - \br_A|_E} 
\nonumber \\ &+ 
\sum_{i=1}^{N_e} \sum_{j=i+1}^{N_e} \frac{Z_A}{|\br_i - \br_j|_E} +
\sum_{A=1}^{N_n} \sum_{B=A+1}^{N_n} \frac{Z_A Z_B}{|\br_A - \br_B|_E} 
\end{align}
where $N_e$ and $N_n$ are the number of electrons and nuclei in the supercell, respectively, $Z$ is the nuclear charge and $|\cdots|_E$ refers to the Euclidean distance, i.e., the distance measured in the embedding space of the Clifford torus.
It is given by
\begin{equation}
\label{Eq:disteuc}
|\br_{\mu} - \br_{\nu}|_E  = 
\frac{1}{\pi} 
\sqrt{ \sum_{s=x}^{x,y,z} L_s^2\sin^2\left[\frac{\pi}{L_s} (s_{\mu} - s_{\nu})\right]}
\end{equation}
where $L_s$ is the length of the supercell along the $s$ axis and the Greek letters refer to electrons or nuclei.
It is implied that the sum is limited to the Cartesian coordinates in which the system is periodic.
For example, if the system is periodic in only one coordinate (arbitrarily chosen to be $x$) the Euclidean distance simplifies to
\begin{equation}
|\br_{\mu} - \br_{\nu}|_E = \frac{L_x}{\pi} \left|\sin\left[\frac{\pi}{L_x} (x_{\mu} - x_{\nu})\right]\right|
\end{equation}
and equals the distance between two points on a ring of circumference $L_x$ (see Fig.~\ref{Distance}).
%

In practice we express the Hamiltonian in Eq.~\eqref{Eq:Hamiltonian} in a basis.
Therefore, we need a basis that is compatible with the imposed periodic boundary conditions.
Moreover, in order to guarantee the accuracy of the results, the basis has to be able 
to describe the sharp oscillating features of the wave function close to the nuclei.
For this purpose we introduce the Clifford gaussian $g^C$.
We define the (unnormalized) one-dimensional Clifford gaussian orbital as
\begin{equation}
g^{C}_i(x) = \left(\frac{L_x}{2\pi} \sin{\left[\frac{2\pi}{L_x}(x-x_A)\right]} \right)^i e^{-\frac{\alpha L_x^2}{\pi^2} \sin^2{[\frac{\pi}{L_x}(x-x_A)]}}
\end{equation}
where $i$ is a non-negative integer.
\begin{figure}[h]
    \centering
    \includegraphics[width=\columnwidth]{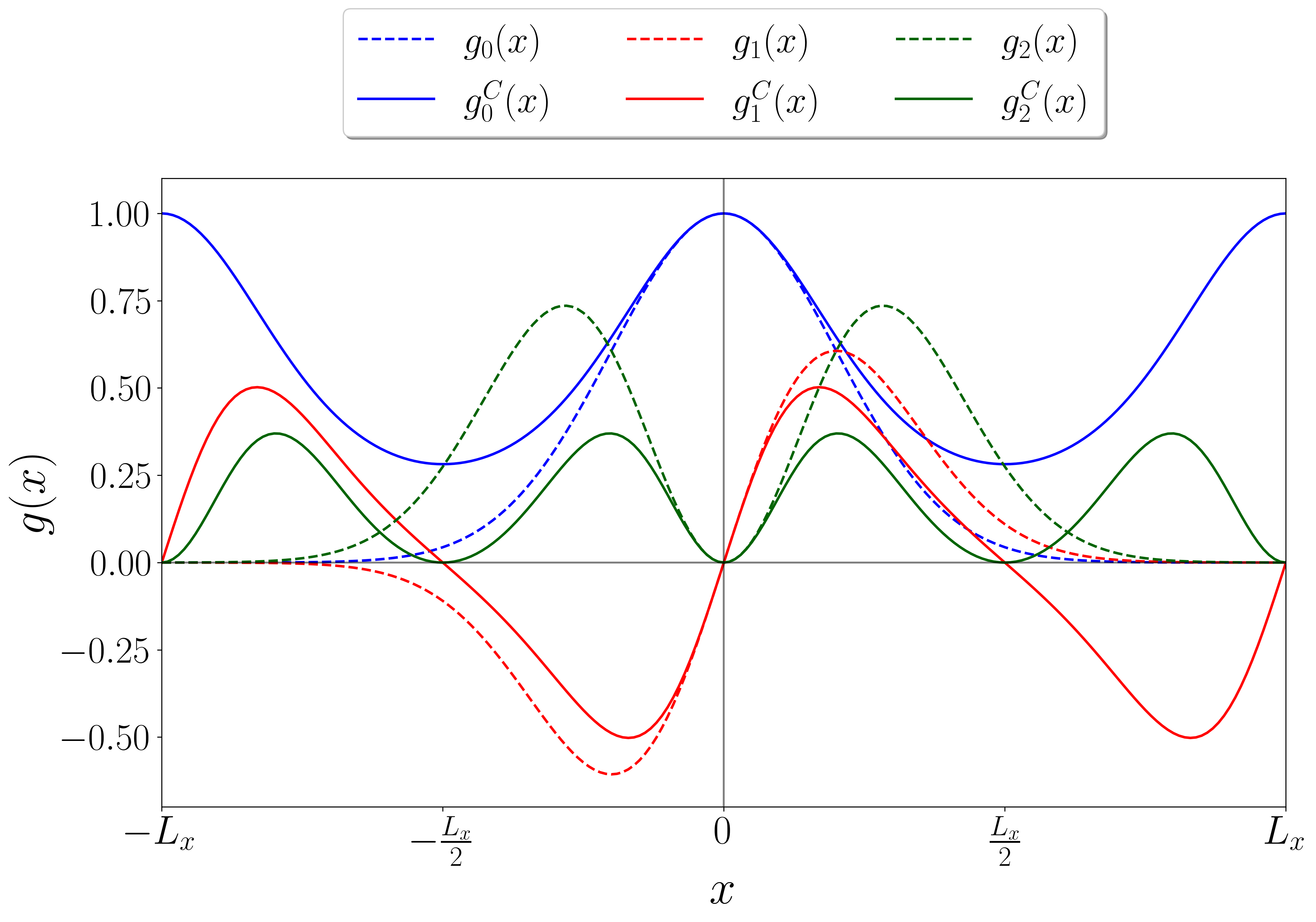}
    \caption{Periodic Clifford gaussians $g^{C}_i(x)$ (solid lines) compared to regular non-periodic gaussians $g_i(x)$ (dashed lines) in two neighboring supercells of length $L_x$ for various values of $i$.}
    \label{Toroidal_gaussian}
\end{figure}
It can be verified that $g^{C}_i(x) = g^{C}_i(x+L_x)$ as it should.
In Fig.~\ref{Toroidal_gaussian} we report Clifford gaussians for several values of $i$ and compare them to the corresponding non-periodic gaussians 
which are given by
\begin{equation}
g_i(x) = (x-x_A)^i e^{-\alpha (x-x_A)^2}
\end{equation}

The total three-dimensional cartesian Clifford gaussian orbital is then given by
\begin{align}
\nonumber
g^{1C}_l(\br) &= g^{C}_i(x) g_j(y) g_k(z)
\\
g^{2C}_l(\br) &= g^{C}_i(x) g^{C}_j(y) g_k(z),
\\
\nonumber
g^{3C}_l(\br) &= g^{C}_i(x) g^{C}_j(y) g^{C}_k(z),
\end{align}
for systems that are periodic in one, two and three Cartesian coordinates, respectively.
Each Clifford gaussian $g^{nC}_l(\br)$ depends on two parameters, the exponent $\alpha$ and the position $\br_A$.
In order to keep the notation simple, in the following it will be implied that in a product of several Clifford gaussians each gaussian depends on a different exponent ($\alpha,\beta,\gamma,\cdots$) and a different position ($\br_A, \br_B, \br_C, \cdots$).

As mentioned in the introduction, the gaussian product rule has been essential to the success of GTO's in quantum-chemistry calculations for atoms and molecules.
The Clifford gaussian function satisfies a similar product rule, i.e., the product of two Clifford gaussian functions also is a Clifford gaussian function~\cite{supmat}.
Thanks to the product rule the one- and two-electron integrals can be reduced to integrals of a single Clifford gaussian.
For example, let us consider the overlap $S_{l_1l_2}$ of two Clifford gaussians,
\begin{equation}
S^{nC}_{l_1l_2} = \int d\br g_{l_1}^{nC}(\br)g_{l_2}^{nC}(\br),
\end{equation}
where $n$ is the number of coordinates in which the system is periodic and it is implied that the upper and lower limits of an integral
with respect to a periodic coordinate $s$ are $0$ and $L_s$, respectively, and $-\infty$ and $\infty$ for a non-periodic coordinate.
In each case the results are purely analytical. 

For example for two $s$-type Clifford gaussians ($l_1=l_2=0$) we obtain
\begin{equation}
S^{nC}_{00} = \left(\frac{\pi}{p}\right)^{\frac{3-n}{2}} 
\prod_{s=x}^{x,y,z} L_s I_0\left(\frac{\gamma_s L_s^2}{2\pi^2}\right) e^{-p\frac{L_s^2}{2\pi^2}},
\end{equation}
where $p=\alpha+\beta$, $I_0$ is a modified Bessel function of the first kind and
\begin{equation}
\gamma_s = \sqrt{\alpha^2+\beta^2 + 2\alpha \beta \cos\left(\frac{2\pi}{L_s}(s_A-s_B)\right)}.
\end{equation}
It is implied that the product is limited to the number of coordinates in which the system is periodic.
We obtain similar analytical expressions for other values of $l_1$ and $l_2$.
It can be readily verified that the Laplacian of a Clifford gaussian also is a Clifford gaussian. 
Therefore, the kinetic-energy integrals can be derived in a similar way as the overlap integrals.
They too are purely analytical.
Thanks to the Laplace transformation we can rewrite the Coulomb interaction in terms of a Clifford gaussian according to
\begin{equation}
\frac{1}{|\br_{\mu}-\br_{\nu}|_E} = \frac{2}{\sqrt{\pi}}\int_0^{\infty} dt e^{-|\br_{\mu}-\br_{\nu}|^2_E t^2}.
\end{equation}
It can be verified that the substitution of the expression for the Euclidean distance in Eq.~\eqref{Eq:disteuc} indeed yields an $s$-type Clifford gaussian.
Therefore, using the product rule for Clifford gaussian functions, the nuclear-attraction integrals as well as the electron-repulsion integrals
can be rewritten as integrals involving a single Clifford gaussian.
The integrals over the cartesian coordinates can then again be performed analytically.
This yields expressions that can be evaluated numerically in an efficient way similarly to the numerical evaluation of the error function (or Boys function) that appears in the nuclear-attraction and electron repulsion integrals when using non-periodic gaussians~\cite{supmat}.
We note that our approach can be extended to other basis sets that have the periodicity of the Clifford supercell.
%

As mentioned in the introduction, a solid that is periodic in $n$ coordinates could, in principle, be represented by an $n$-torus but this is either cumbersome (1D and 2D) or impossible (3D).
Here we show that our approach based on Clifford periodic boundary conditions and Clifford gaussians becomes equivalent to this representation in the thermodynamic limit while being numerically feasible for all dimensions.
We demonstrate this equivalence by calculating the ground-state energy of periodic hydrogen chains, described as a ring with regular non-periodic gaussians and as a quasi-one-dimensional Clifford torus with periodic Clifford gaussians.
We expect that the ring and the quasi-one-dimensional Clifford torus become equivalent in the thermodynamic limit 
because the transverse component of the electron density becomes infinitesimally small with respect to its longitudinal component in this limit.
We used an interatomic distance of 1.8 a.u., which is close to the equilibrium distance of the infinite chain, and the pob-tzpv basis set~\cite{peintinger2013consistent} which was specifically developed for solid-state calculations.
The computer code with which all the one- and two-electron integrals were calculated can be found in Ref.~\cite{TGaussian_code}.

\begin{figure}[t]
    \centering
    \includegraphics[width=\linewidth]{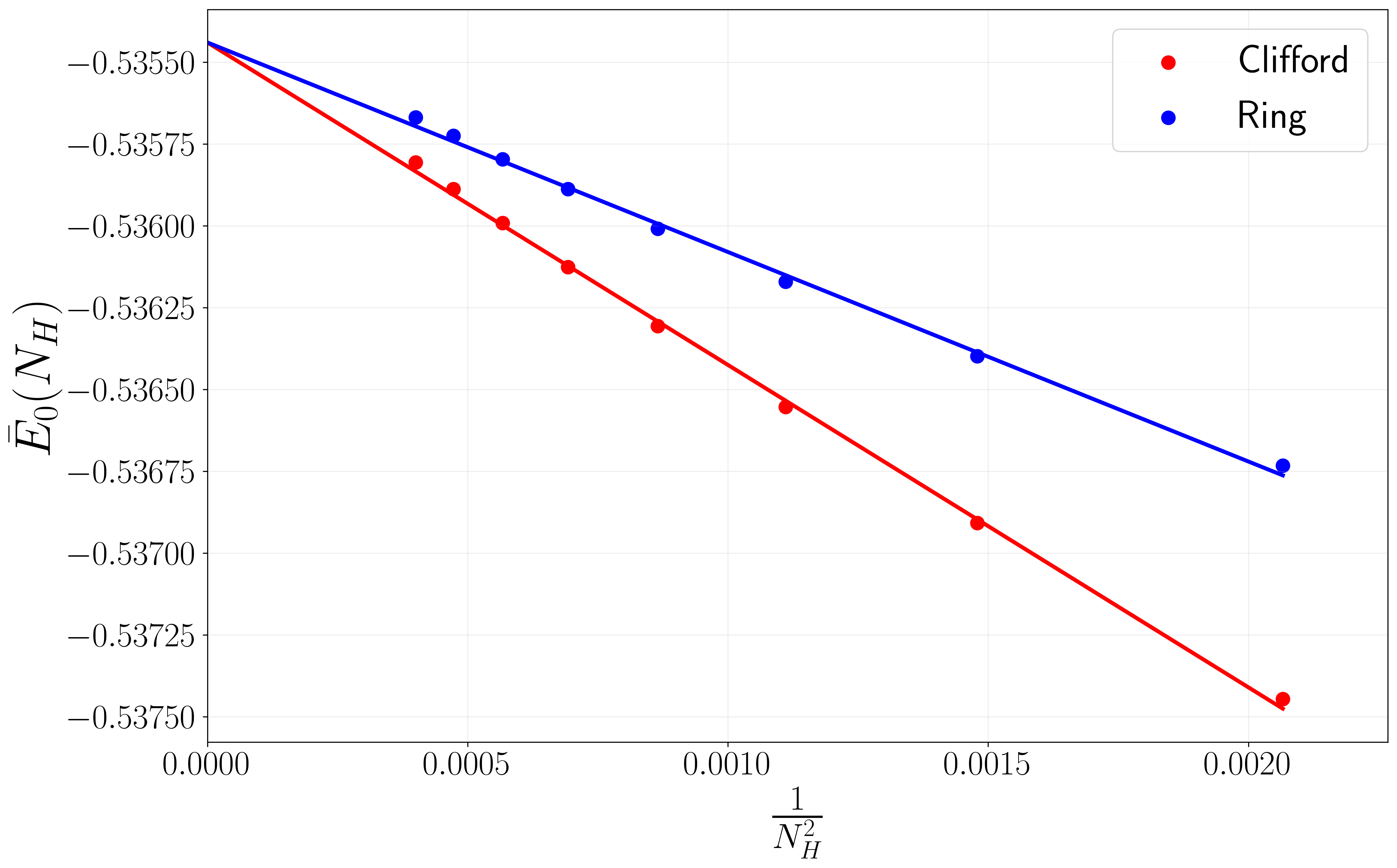}
    \caption{Hartree-Fock ground-state energies per atom $\bar{E}_0(H_N) $ of hydrogen chains with various numbers of atoms and a nearest-neighbor distance of 1.8 a.u. using Clifford periodic boundary conditions and Clifford gaussians (red dots) and a ring configuration with non-periodic gaussians (blue dots).
    The corresponding solid lines represent the extrapolation to the thermodynamic limit according to Eqs.~\eqref{Eq:TDL_torus} and \eqref{Eq:TDL_ring}, respectively.}
    \label{pobtzpv}
\end{figure}

In Fig.~\ref{pobtzpv} we report the Hartree-Fock ground-state energies (per hydrogen atom) of periodic hydrogen chains as a function of $N_H^{-2}$ where $N_H$ is the number of hydrogens in the chain.
We see that the ground-state energies per atom $\bar{E}_0(N_H)=E_0(N_H)/N_H$ corresponding to both the ring and the Clifford torus are almost straight lines indicating that the energies converge as $N_H^{-2}$ contrary to ground-state energies obtained within open-boundary conditions which converge at the slower rate of $N_H^{-1}$~\cite{Motta_2017}.
We can now extrapolate the energies per atom in Fig.~\ref{pobtzpv} to the thermodynamic limit according to
\begin{equation}
\bar{E}_{0}(N_H)= \bar{E}_{0}^{TDL} - \frac{a}{N_H^2}
\end{equation}
where $\bar{E}_{0}^{TDL}$ and $a$ are constants with the former corresponding to the ground-state energy per atom in the thermodynamic limit (TDL).
A mean-squared error fit yields the following expressions for $\bar{E}(N_H)$,
\begin{align}
\bar{E}_0(N_H) &= -0.53544 - \frac{0.98494}{N_H^2} \quad \text{(Clifford torus)}
\label{Eq:TDL_torus}
\\
\bar{E}_0(N_H) &= -0.53544 - \frac{0.64002}{N_H^2}\quad \text{(ring)}
\label{Eq:TDL_ring}
\end{align}
for the Clifford torus and the ring, respectively.
Remarkably, the two approaches yield the exact same value in the thermodynamic limit, demonstrating the numerical equivalence of the Clifford torus and the ring in the thermodynamic limit.
We note that this result is independent of the hydrogen chains we include in the fit.
Although changing the chains used in the fit will slightly change $\bar{E}_{0}^{TDL}$, it will yield the same value for the ring and Clifford torus.

The main advantage of our approach is that the required modifications to an existing implementation of a quantum-chemistry method for non-periodic systems such as atoms and molecules only involve the one- and two-electron integrals.
These integrals only have to be calculated once and are independent of the quantum-chemistry method.
Therefore, our approach can be seamlessly combined with any quantum-chemistry method without any further modifications to existing implementations.
To illustrate this we have read the integrals calculated within our approach into Quantum Package~\cite{Garniron_2019}, which is a quantum-chemistry code for atoms and molecules.
We were thus able to perform periodic coupled-cluster calculations of the hydrogen chains.

\begin{figure}[t]
    \centering
    \includegraphics[width=\linewidth]{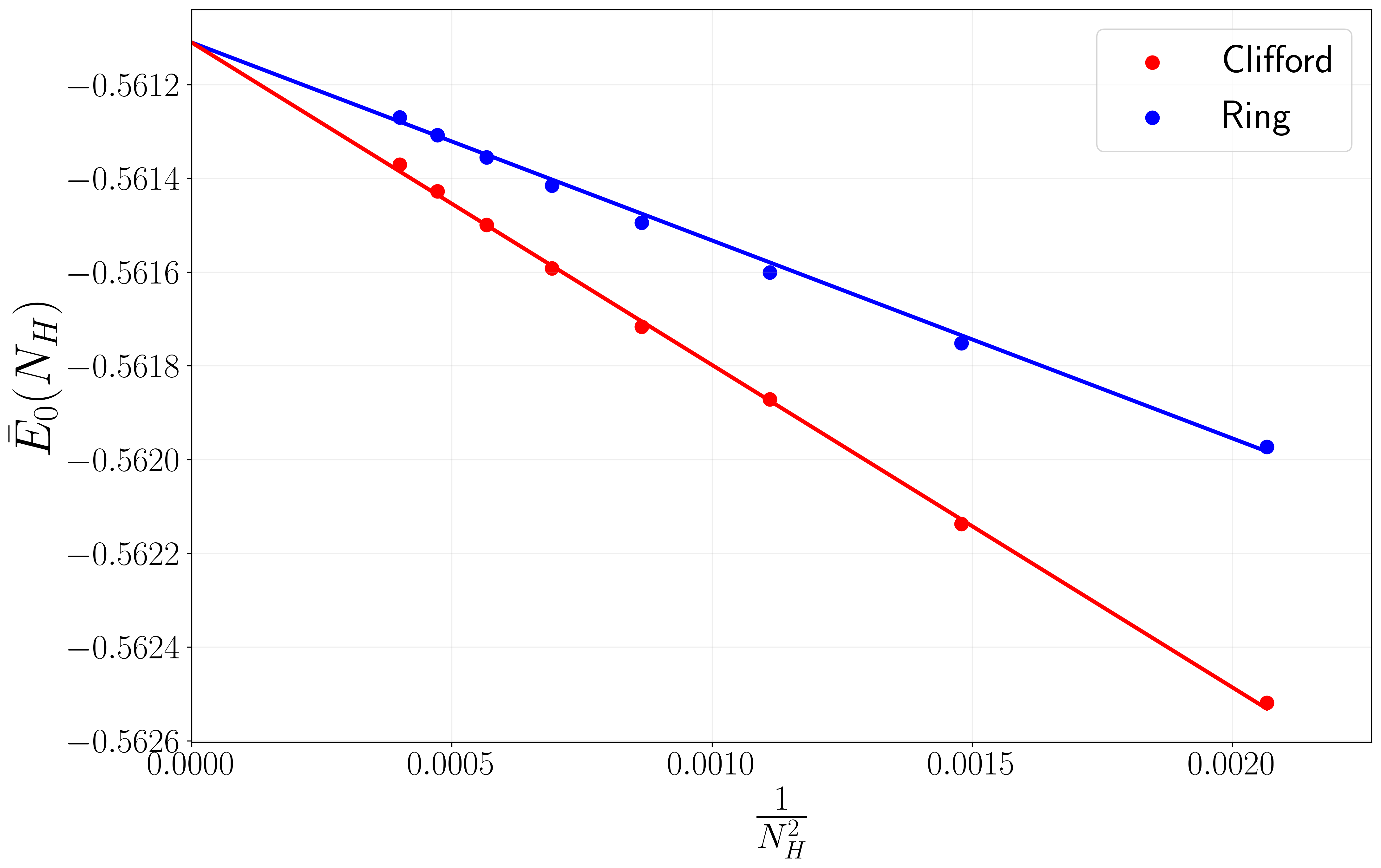}
    \caption{CCSD(T) ground-state energies per atom $\bar{E}_0(H_N) $ of hydrogen chains with various numbers of atoms and a nearest-neighbor distance of 1.8 a.u. using Clifford periodic boundary conditions and Clifford gaussians (red dots) and a ring configuration with non-periodic gaussians (blue dots).
    The corresponding solid lines represent the extrapolation to the thermodynamic limit according to Eqs.~\eqref{Eq:TDL_torus_CC} and \eqref{Eq:TDL_ring_CC}, respectively.}
    \label{CCSDT}
\end{figure}

The gold-standard in quantum chemistry to account for electron correlation is the coupled-cluster method with singles, doubles and perturbative triples (CCSD(T))~\cite{Cizek_1966}.
Therefore, in Fig.~\ref{CCSDT} we report the CCSD(T) ground-state energies per hydrogen atom of the hydrogen chains obtained within our approach using a periodic Clifford torus and Clifford gaussians.
We compare these energies to the corresponding energies of the ring obtained using non-periodic gaussians.
As was the case for Hartree-Fock, the results are almost linear as a function of $N_H^{-2}$.
We extrapolate the energies per atom in Fig.~\ref{CCSDT} to the thermodynamic limit and we obtain
\begin{align}
\bar{E}_0(N_H) &= -0.56111 - \frac{0.68804}{N_H^2} \quad \text{(Clifford torus)}
\label{Eq:TDL_torus_CC}
\\
\bar{E}_0(N_H) &= -0.56111 - \frac{0.42239}{N_H^2} \quad \text{(ring)}
\label{Eq:TDL_ring_CC}
\end{align}
Also for CCSD(T) the two approaches yield the same energy in the thermodynamic limit, again showcasing the equivalence of the Clifford torus and the ring in the thermodynamic limit.
%

In conclusion, we have presented a general formalism to calculate solid-state properties using quantum-chemistry methods.
In our approach we represent a periodic solid as a supercell that has the topology of a Clifford torus which is both periodic and flat.
Practical calculations are then made feasible thanks to a periodic basis set of Clifford gaussians
which are periodic functions that are naturally adapted to the topology of the Clifford torus.
We recover the properties of the solid by extrapolating the results obtained for Clifford supercells of various sizes to the thermodynamic limit.
Our approach can treat systems that are periodic in 1, 2 and 3 dimensions.
We illustrated our approach by applying it to the ground-state energy of a periodic chain of hydrogen atoms for which we can compare our results to those of a ring of hydrogen atoms.
We showed that both approaches are numerically equivalent in the thermodynamic limit.
Finally, we demonstrated that our approach can be seamlessly combined with existing implementations of quantum-chemistry methods in computer codes for atoms and molecules.
These codes can thus be immediately used to calculate the ground- and excited-state properties of periodic solids.
\section*{Acknowledgment}
We thank the French Agence Nationale de la Recherche (ANR) for financial support (Grant Agreement ANR-22-CE29-0001).
\end{document}